\documentclass[useAMS,usenatbib]{mn2e}
\usepackage{times}
\usepackage{amsmath}
\usepackage{amssymb,amsfonts}
\usepackage{graphicx}
\usepackage{verbatim}
\usepackage{psfig}
\usepackage{epsf}
\title
[The INT/WFC survey of the Monoceros Ring: Accretion origin or
  Galactic Anomaly?]
{The INT/WFC survey of the Monoceros Ring: Accretion origin or
  Galactic Anomaly?}
\author[Blair Conn et al.]
       {Blair C. Conn,$^1$ Geraint F. Lewis,$^1$ Mike J. Irwin,$^2$ Rodrigo A. Ibata,$^3$
	 \newauthor Annette M. N. Ferguson,$^4$ Nial Tanvir$^5$ \& Jonathan M. Irwin$^2$\\
$^{1}$Institute of Astronomy, School of Physics, A29, University of Sydney, NSW 2006, Australia:\\
Email \tt{bconn@physics.usyd.edu.au}, \tt{gfl@physics.usyd.edu.au}\\
$^{2}$Institute of Astronomy, Madingley Road, Cambridge, CB3 0HA, U.K.: Email \tt{mike@ast.cam.ac.uk},\tt{jmi@ast.cam.ac.uk}\\
$^{3}$Observatoire de Strabourg, 11, rue de l'Universit\'e, F-67000, Strasbourg,
France: Email \tt{ibata@astro.u-strasbg.fr}\\
$^{4}$Institute for Astronomy, University of Edinburgh, Royal Observatory, Blackford Hill, Edinburgh, EH9 3HJ, U.K.:Email \tt{ferguson@roe.ac.uk}\\
$^{5}$Centre for Astrophysics Research, University of Hertfordshire,
College Lane, Hatfield AL10 9AB U.K.: Email ~\tt{nrt@star.herts.ac.uk}}

\begin{document}

\date{\today \hspace{10pt}(Version 1.4)} 

\pagerange{\pageref{firstpage}--\pageref{lastpage}} \pubyear{2005}

\def\LaTeX{L\kern-.36em\raise.3ex\hbox{a}\kern-.15em
    T\kern-.1667em\lower.7ex\hbox{E}\kern-.125emX}

\newtheorem{theorem}{Theorem}[section]

\label{firstpage}

\maketitle

\begin{abstract}
We present the results of a wide-field camera survey of the stars in
the Monoceros Ring, thought to be an additional structure in the Milky
Way of unknown origin. Lying roughly in the plane of the Milky Way,
this may represent a unique equatorial accretion event which is
contributing to the Thick Disk of the Galaxy. Alternatively, the
Monoceros Ring may be a natural part of the Disk formation process.
With ten pointings in symmetric pairs above and below the plane of the
Galaxy, this survey spans 90 degrees about the Milky Way's
equator. Signatures of the stream of stars were detected in three
fields, ({\it l},{\it b}) = (118,+16)$^\circ$ and
(150,+15)$^\circ$ plus a more tentative detection at (150,-15)$^\circ$.
Galactocentric distance estimates to these structures gave
$\sim$17, $\sim$17, and $\sim$13 kpc
respectively.  The Monoceros Ring seems to be present on both sides of
the Galactic plane, in a form different to that of the Galactic warp,
suggestive of a tidal origin with streams multiply wrapping the
Galaxy. A new model of the stream has shown a strong coincidence with
our results and has also provided the opportunity to make several more
detections in fields in which the stream is less significant. The
confirmed detection at ({\it l},{\it b}) = (123,-19)$^\circ$ at
$\sim$14, kpc from the Galactic centre allows a
re-examination revealing a tentative new detection with a
Galactocentric distance of $\sim$21 kpc. These detections
also lie very close to the newly discovered structure in
Triangulum-Andromedae hinting of a link between the two. The remaining
six fields are apparently non-detections although in light of these
new models, closer inspection reveals tentative structure. With the
overdensity of M-giant stars in Canis Major being claimed both as a
progenitor to the Monoceros Ring and alternatively a manifestation of
the Milky Way warp, much is still unknown about this structure and its
connection to the Monoceros Ring. Further constraints are needed for
the numerical simulations to adequately resolve the increasingly
complex view of this structure.
\end{abstract}

\begin{keywords}
Galaxy:\hspace{2pt}formation -- Galaxy:\hspace{2pt}structure -- galaxies:\hspace{2pt}interactions
\end{keywords}

\section{Introduction}
The formation and evolution of galaxies remains one of the big
questions in astronomy. In the currently favoured $\Lambda$CDM model,
galaxies are built up over time via the accretion of smaller systems
\citep[e.g.][]{1978ApJ...225..357S,1978MNRAS.183..341W,1978MNRAS.184..185W,2003ApJ...597...21A,2003ApJ...591..499A}.
The picture is not wholly satisfactory and some parts of the Milky Way
may have formed in an initial collapse of baryonic material, somewhat
akin to the model of galaxy formation proposed by \citet{1962ApJ...136..748E}.

One firm prediction of the $\Lambda$CDM model is that this accretion
of smaller systems should still be ongoing and that the Milky Way Halo
should contain a large number of satellite systems.  It has been
suggested that, given the model, there are too few satellites actually
within the Milky Way Halo~\citep{1999ApJ...522...82K}, although the
extent of this discrepancy has been a matter of some debate. Many
attempts have been made to resolve this issue
\citep[e.g.][]{2001AIPC..586...73M} by correcting the measured velocity
dispersion of the local satellites to a velocity dispersion of the
dark matter components, while more recently \citet{2004ApJ...610L...5S} used
numerical methods to highlight a scenario whereby these low mass
systems are evaporated at the Epoch of Re-ionization.  In the
meantime, more and more surveys are probing the Halo of the Milky Way
revealing structures which may point to the formation history of our
galaxy \citep[see][]{2002ARA&A..40..487F}.

The tidal dismemberment of a dwarf galaxy as it falls through the
Milky Way Halo is a slow process, with extensive streams of tidal
debris existing for long periods of time \citep{1998ApJ...500..575I,
1999ApJ...512L.109J}. While ancient remnants have been identified, via
phase-space analysis, in our own Galactic neighbourhood
\citep{1999Natur.402...53H,2003ApJ...585L.125B}, more extensive
surveys of the Galactic Halo, such as the Spaghetti Survey
\citep{2000AJ....119.2254M} and utilizing 2MASS
\citep{2003ApJ...599.1082M} have concluded that there is only a
single, major on-going accretion event, that of the Sagittarius dwarf
galaxy \citep{1994Natur.370..194I}.  While this accretion event is
adding mass to the Galactic Halo and provides an important probe of
the shape of the dark matter potential
\citep[e.g.][]{2001ApJ...551..294I,2004AJ....128..245M}, the lack of other other major
accretion events is somewhat disconcerting given the predictions from
$\Lambda$CDM.

As will be discussed in detail in $\S$\ref{ring}, the recently
discovered Monoceros Ring (MRi) can be interpreted as an additional
on-going accretion event within the Milky Way. Investigating the
density and extent of this structure is important when trying to fully
understand the impact this type of event is having on the evolution of our
galaxy both in the past and into the future. If the MRi is instead the
outermost edge of the Milky Way, mapping the outer reaches of the Disk will provide
insight into the Milky Way's past.  To this end, we have used
the Isaac Newton Telescope Wide Field Camera to continue a campaign to
detect this stellar population around the Galactic plane mapping out
the extent of the MRi.  This paper presents the results of a
wide-field camera survey of the extensive stellar population thought
to represent a continuation of the previous work in this field. The layout of the paper is as
follows; $\S$\ref{ring} summarises the extant knowledge of the MRi and
the associated population of stars while $\S$\ref{obs} describes the
observational procedure and data reduction. $\S$\ref{analysis}
describes the analysis procedure, and the conclusions of this study
are presented in $\S$\ref{conclusions}.

\begin{figure*}
\centerline{ \psfig{figure=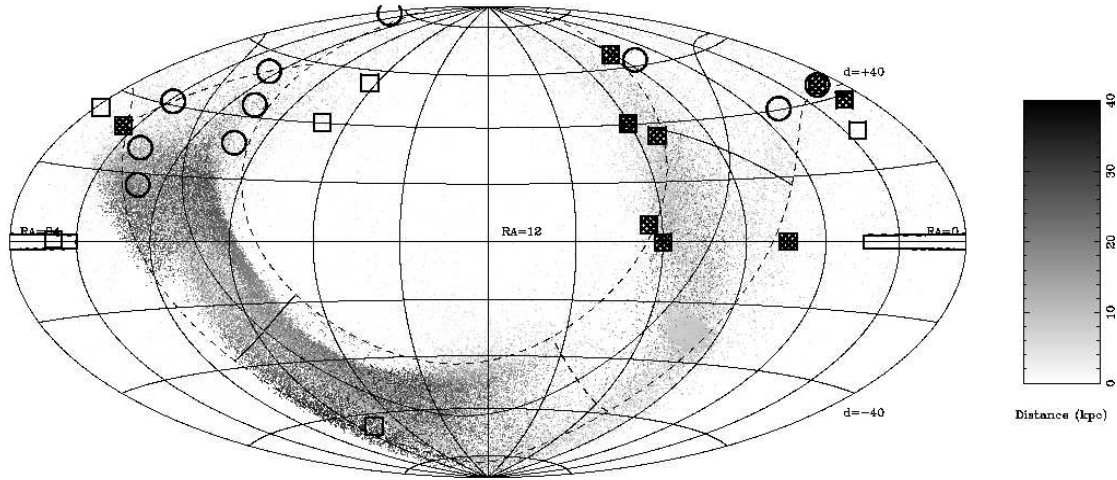,angle=270,width=5.8in}}
\caption[]{Aitoff projection of the sky, illustrating the locations of
the fields obtained for this current survey.  The projection is in
equatorial coordinates, with the Milky Way Equator ({\it b} = 0$^\circ$)
shown as a solid black curve accompanied by the dashed curves which
mark out Galactic latitudes of $b = \pm 20^\circ$. The Galactic centre
({\it l} = 0$^\circ$), and anti-centre({\it l} = 180$^\circ$) are shown as
solid bars crossing the Galactic equator. The points
represent the prograde model for the destruction of the Canis Major
dwarf, taken from \citet{2004MNRAS.348...12M} with the grayscale showing
the Galactocentric distance (kpc) of the points, as shown in the
sidebar. The dense knot located below the Galactic equator near RA
$\sim$7.5 hrs and Dec $\sim$-30 degrees, represents the final location
of the progenitor used in the simulation. The symbols in the plot are
represented as follows: the circles represent the location of
the fields in this survey; the squares represent the fields associated
with the M31 survey ~\citep{2001Natur.412...49I,2002AJ....124.1452F},
with the hashed squares being confirmed detections of the ring, with
empty squares being non-detections.
\label{figaitoff}}
\end{figure*}

\section{The Monoceros Ring}\label{ring}

The first sign of a new structure in the Galaxy, the Monoceros
Ring\footnote{The stream of stars  at  the edge  of  the Milky  Way
  has  received several  names, including  The  One
  Ring~\citep{2003MNRAS.340L..21I},  the  Monoceros Stream
  \citep{2003ApJ...588..824Y}  and  the   Galactic  Anti-Centre
Stellar Stream  -- GASS -- \citep{2003ApJ...599.1082M}.   For the sake
of  simplicity,  throughout  this  paper this  stellar  population  is
referred to as  the Monoceros Ring (MRi),  although this name does
not  imply  we  currently  know  the  exact  nature  of  this  stellar
population.}, came  from a  study of colour selected F-stars  drawn from
commissioning    data   from    the   Sloan    Digital    Sky   Survey
[SDSS,~\citet{2002ApJ...569..245N}]. Obtained in  a narrow strip around
the celestial  equator these revealed two  major stellar overdensities
in the Galactic Halo, consistent with an intersection of the streams of
tidal     debris      torn     from     the      Sagittarius     Dwarf
galaxy~\citep{2001ApJ...547L.133I}.  Accompanying  these, however, was
an additional significant  overdensity in  the  direction  of the  Galactic
anti-centre,  interpreted  as being  another  tidal  stream about  the
Galaxy, just past the edge of the stellar Disk, at a Galactocentric
distance of $\sim18$kpc and with a thickness of $\sim2$kpc, covering
over  $\sim50^\circ$ of sky within  $|b|<30^\circ$  of  the Galactic  equator.

Given the results of the study of the SDSS, \citet{2003MNRAS.340L..21I} searched for
the signature  of this stellar  population in fields obtained  for the
Isaac Newton Telescope Wide Field Camera (INT/WFC) survey ~\citep[see][]{2001NewAR..45...97M}. Identifying
the  stream  as a  distinct  population  in colour-magnitude  diagrams
(CMDs), this study found the stars to be extensively distributed; over
100$^\circ$ of the sky within  ${\rm |b|<30^o}$ of the Galactic equator,
suggesting that the stream  completely rings the Galaxy. Main sequence
fitting  reveals  that  the  Galactocentric distance  to  the  stream
varies  between   $\sim15$kpc  and  $\sim20$kpc,   with  an  apparent
scale-height of  $\sim0.75$kpc.  \citet{2003MNRAS.340L..21I} suggested
that  this  stream represents  debris  of  an  accreting dwarf  galaxy,
although  pointed  out   that  the  extant  data  did   not  rule  out
alternatives such as an outer  spiral arm or unknown warp/flare of the
Galactic Disk. In fact, they favoured the interpretation of it being a
perturbation of the disc, possibly the result of ancient warps.

Simultaneously,  \citet{2003ApJ...588..824Y} presented an  analysis of
a  larger SDSS  catalog  of Halo  stars obtaining
a  number of  radial velocities  in  several fields  over the  stream,
finding  a velocity dispersion  of $\sim25-30$km/s,  inconsistent with
any known  Galactic  component.  Furthermore,  these  kinematics
indicate that  the stream possesses  a prograde orbit about  the Milky
Way  with  a  circular  velocity of  $\sim215\pm25$km/s [Note: this revised
value was presented in \citet{2004ApJ...605..575Y}].   Metallicity
estimates from  these spectra indicate the stars  are relatively metal
poor    $([\frac{Fe}{H}]\sim-1.6)$.    \citet{2003ApJ...588..824Y}   also
concluded the stream represents a cannibalized dwarf galaxy undulating
about the edge of the Galactic Disk.

Several other  studies have focused  upon this  stream population;
using the 2-Micron  All Sky Survey (2MASS) \citet{2003ApJ...594L.115R}
identified M-giant stars beyond the  Disk of the Milky Way, consistent
with the population  detected in the optical. Again,  at a
Galactocentric distance (R$_{GC}$) of
$\sim18\pm2$kpc, this  arc of stars covers $\sim170^\circ$,  with a higher
metallicity than  previously determined $([\frac{Fe}{H}]=-0.4\pm0.3)$.
~\citet{2003ApJ...594L.119C}  extended this  study,  obtaining stellar
velocities  of M-giants  selected from  2MASS.  Confirming  a velocity
dispersion of $\sim20$km/s, this  study concluded the stream orbits the
Milky  Way  in  a  prograde  fashion  on an  orbit  with  very  little
eccentricity. While  this may seem  problematic for accretion models
preferring elliptical orbits, numerical simulations
by \citet{2003ApJ...592L..25H} suggest that tidal streams in the plane
of the  Milky Way can  possess quite circular orbits;  these numerical
studies, however, also suggest that the accretion event must be young,
less  than $\sim1$Gyr  since its  first perigalactic  passage,  or any
coherent     structure      would     have     dissolved.      Finally,
\citet{2004ApJ...602L..21F}  noted five  globular  clusters which  are
apparently  associated with the  stellar stream,  as well  as $\sim15$
outer,  old   stellar  clusters,  bolstering  the   argument  that  it
represents an accreting dwarf galaxy.

\citet{2004MNRAS.348...12M} also employed the 2MASS catalogue to
identify M-giants beyond the Disk of the Milky Way. By considering the
projected density of these stars, this study uncovered
north-south anisotropies in the density of M-stars, with 
arcs above and below the plane of the Galaxy.  Significantly,
\citet{2004MNRAS.348...12M} identified a strong overdensity of these stars,
$\sim2300$ of them in roughly 20$^\circ$ of the sky at
$(l,b)\sim(240,-8)^\circ$, the constellation of Canis Major (CMa). This
is a similar number of M-stars to that seen in the Sagittarius dwarf
galaxy and \citet{2004MNRAS.348...12M} similarly interpret this
population of stars as a dwarf galaxy. It is probably the progenitor,
with an original mass of $\sim10^8-10^9{\rm M_\odot}$, of the extensive
stream of stars around the edge of the Disk of the Milky Way. Given
its mass, if CMa does represent an equatorial accretion event,
it will, when fully dissolved, increase the mass of the Thick Disk by
$\sim10\%$.

Additional evidence for this accretion interpretation of the CMa dwarf
comes from \cite{2004MNRAS.354.1263B} who found the signature of the main
body of the dwarf in the background to several Galactic open clusters,
with the analysis of this population suggesting that it is somewhat
metal rich with an age of $\sim2-7$ Gyrs, although a blue plume
indicates a younger population.  This study also identified several
globular and open clusters associated with the dwarf.  Recently,
~\citet{2004AJ....127.3394F} have shown that the globular clusters associated
with CMa possess their own age-metallicity relationship which is
distinct from that of the Galactic population.  Furthermore, the
clusters are smaller than expected, if drawn from the Galactic
population, strengthening the interpretation that they are of non-Galactic origin.

\cite{2004A&A...421L..29M}, by re-analysing the 2MASS data claim the overabundance of M-giants in
CMa is simply a signature of the warp in the Milky Way.  In response, ~\citet{2004MNRAS.355L..33M} used 2dF\footnote{The 2dF
  instrument on the Anglo-Australian Telescope is a multi-fibre
  spectrograph with the ability to obtain spectra for 400 stars in a
  single pointing, inside
a two-degree field.} data of the Canis
Major region to highlight the velocity disparities between the Milky
Way Disk stars and the M-giant overdensity stars.  This has been
repeated more recently by ~\citet{2005Martin} suggesting the dwarf galaxy
CMa can now be tentatively accepted as a real entity despite the
current debate.  Connecting CMa
to the MRi is more problematic due to gaps in the detection of the
ring and poor kinematic knowledge.  A recently completed 2dF kinematic
survey~\citep{2005Martin} of the Canis Major region may yield results in this area
although currently any general conclusions linking it with the MRi are speculative.

More information about Halo substructure has been recently discovered
as \cite{2004ApJ...615..732R} have identified another structure in
Triangulum-Andromedae (TriAnd) which extends much further out than the
MRi.  Currently it is not known whether the two are related, although
interestingly the detection of the MRi by \cite{2003MNRAS.340L..21I} resides at the edge
of this new structure in TriAnd suggesting a connection.

The latest evidence supporting the interpretation of the Canis Major Dwarf
galaxy has been presented by \cite{Martinez2004}, showing the
colour-magnitude diagram of a region (0.5$^\circ$ x 0.5$^\circ$) centred on
({\it l},{\it b}) = (240,-8)$^\circ$, the overdensity proposed by
\cite{2004MNRAS.355L..33M}.  An upper limit to the Heliocentric distance of the
galaxy is found to be 5.3$\pm$0.2 to 8.1$\pm$0.4 kpc; this is comparable with
previous estimates. By measuring the surface brightness and total
luminosity of the dwarf galaxy an estimate of the mass range is found
to be 1.0$\times 10^8 <$ M$_\odot$ $<$ 5.5$\times$10$^8$.  The tightness of the  main sequence
they found contradicts the claims that the Canis Major overdensity is
the signature of the Galactic warp. \cite{Pena2004} have completed
extensive modelling of the MRi and while not conclusively establishing
a connection between Canis Major and the MRi, their results are highly
suggestive of such a link.  \cite{Pena2004} using all of the current
information known about the MRi, undertook thousands of simulations,
prograde and retrograde, in an attempt to find a model which best
fitted the data. While some retrograde models were marginally
acceptable, their preferred model was a prograde orbit which includes
multiple wraps of the Milky Way.  This allows for the scenario that
the tidal stream has both near and far components. $\S$\ref{123m19}
discusses a field in which this phenomenon appears. If the dwarf galaxy
has completed more than one orbit then the stream must be a much older
structure than previously assumed [cf. \cite{2003ApJ...592L..25H}], raising the
possibility that the newly discovered TriAnd structure
\citep{2004ApJ...615..732R} could be the distant arm of a multiply
wrapped tidal stream.  

The model of \cite{Pena2004} and the continuing work of those
studying this new Galactic feature are slowly piecing together this
structure, although currently there is no real coverage around the
entire Galactic plane, these observations hopefully extend the
knowledge of the MRi and it's potential progenitor the Canis Major
Dwarf.

\section{Observations and Reduction}\label{obs}
The data was obtained on  the Isaac Newton Telescope Wide Field Camera (INT/WFC)
at Roque de Los Muchachos in  La Palma, Canary Islands. Mounted at the telescope
prime  focus, this  covers  0.29 square  degrees  field per  pointing,
imaging  onto  four  4k$\times$2k  CCDs.  These  possess  13.5$\mu$m  pixels,
corresponding to a pixel scale of 0.33 arcsec per pixel.

Nine survey regions were  chosen, roughly equally spaced between {\it
l} =  $61^o -  150^\circ$. To aid in determining the location of the
fields, the model of ~\citet{2004MNRAS.348...12M} (See
Figure~\ref{figaitoff}) was used to predict where we might expect the
spread of debris from an equatorial accretion. For each Galactic longitude, two regions
symmetrically placed above and  below the Galactic plane, were imaged.
Each target region is a composite of overlapping fields, the number of
which depended on time available to observe in each region (details of
the   observations  are   summarised  in   Table~\ref{ObsTable} and
presented graphically in Figure~\ref{figaitoff}),  but
typically with a total  area of $\sim2$ square degrees.  With the representative
integration times, the limiting magnitudes are on average for these
observations $V_\circ \sim$ 23.3, $i_\circ \sim$ 22.2, $g_\circ \sim$ 23.8 and
$r_\circ \sim$ 22.8.  Archival data of (123,-19)$^\circ$, taken from the M31 survey
\citep{2001Natur.412...49I,2002AJ....124.1452F}, is used as the
opposing field to (118,+16)$^\circ$.

De-biassing and  trimming, vignetting correction, astrometry  and photometry were
all    undertaken   with    the   CASU    data    reduction   pipeline
~\citep{2001NewAR..45..105I}.   The flat  fielding  employed a  master
twilight flat generated  over the entire observing run.  Each star was
individually extinction corrected using the dust\_getval.c program supplied by
Schlegel\footnote{http://www.astro.princeton.edu/$\sim$schlegel/dust/data/data.html}
which interpolates the extinction using the maps presented by
~\citet{1998ApJ...500..525S}. Observing several standard fields per
night allows the calibration of the photometry to be determined, as
described by ~\citep{2001NewAR..45..105I}, deriving the CCD
zero-points.  These are consistent to within
a few percent on photometric nights.  The data reduction pipeline produces
a catalogue  of all  images in each colour-band. Rejecting non-stellar
images, the catalogues can be cross-correlated  and   the  colour
for  individual  stars   can  be determined. Near the limiting
magnitude, however, galaxies can appear stellar and so galactic contamination
is expected for the faintest sources. 

\begin{table*}
\centering
\caption{Summary of the observations  of Canis Major Tidal Stream with
  the INT/WFC, ordered in ascending Galactic longitude ({\it l}). }
\begin{tabular}{rcccccccl} \hline \hline
Fields\it{(l,b)$^\circ$} & No. of Regions per field & Exposure (s) &
Filter & Average Seeing (arcsec) & Total Area ($deg^2$) & Detection &
 Date \\ \hline
 (61,$\pm$15)$^\circ$  & 6  & -         & $g$, $r$ & -   & 1.74    & No (+/-)& late 05/03        \\
 (75,$\pm$15)$^\circ$  & 7 & 600s      & $V$, $i$ & 1.3 & 2.03  &  No (+/-)&  05/09/03\\
 (90,$\pm$10)$^\circ$  & 7 & 900s      & $g$, $r$ & 1.0 & 2.03  & No (+/-)&  28-31/08/03\\
(118,$+$16)$^\circ$     & 11 & 900/1000s & $V$, $i$ & 1.2 & 3.19 & Yes &  29/08-02/09/03\\
(123,$-$19)$^\circ$     & -  & -         & $V$, $i$ & -   & -    & Yes & -        \\
(150,$\pm$15)$^\circ$ & 6 & 600s      & $V$, $i$ & 1.3 & 1.74 & Yes(+), Maybe(-) & 03-05/09/03\\
\hline\hline
\end{tabular}
\label{ObsTable}
\end{table*}

\section{Analysis}\label{analysis}
\subsection{Detecting the Monoceros Ring}\label{detections}

\begin{figure}
\centerline{ \psfig{figure=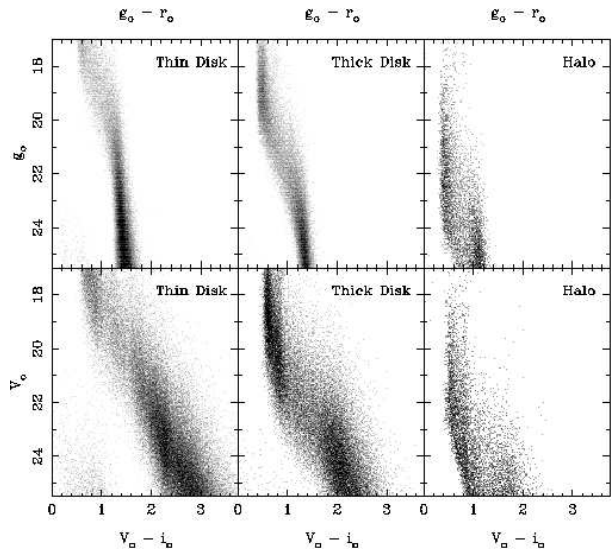,angle=0,width=8cm}}
\caption[]{The top panels show the Hess plots of (61,-15)$^\circ$ field from the
  synthetic model of the galaxy by \citet{2003A&A...409..523R} being split into its various
  Galactic components' Thin Disk, Thick Disk and Halo.  For $g_\circ$ and
  $r_\circ$ CMDs, this plot illustrates where the various components of
  the galaxy lie on the CMD.  The lower panels are from the model
  field (75,+15)$^\circ$, and show how a $V_\circ$,$i_\circ$ CMD can be similarly
  deconstructed.
\label{figmodel}}
\end{figure}

As a first step each field was investigated by eye, looking for
the presence of the MRi main sequence.  In previous studies
\citep{2003MNRAS.340L..21I} the MRi structure was observed as a
coherent stellar population superimposed upon an overall Galactic CMD.
Hence, we compare the observed and synthetic CMDs, as well as the
North and South fields, to search for such a signal.

The synthetic CMDs were generated from the online models of the Milky
Way by \citet{2003A&A...409..523R}\footnote{http://www.obs-besancon.fr/www/modele}.
To avoid any artificial cuts in the data, the modelled fields are
extended out to 50 kpc (Heliocentric distance, R$_{HC}$) and include
the entire magnitude range available to the model (-99,99) for each
passband. The extinction parameter is set to zero, all ages/populations
and luminosity classes are included.  This creates a complete picture
of the field of interest and allows any later cuts on magnitude to be
made at our discretion.  The ($V$,$i$) figures have been created using data
straight from the model while the ($g$,$r$) figures converted the model
($V$,$i$) magnitudes via the INT/WFC colour
corrections\footnote{http://www.ast.cam.ac.uk/$\sim$wfcsur/technical/photom/colours}.

To understand the CMDs, requires that we know what Galactic features
are present in the data. By deconstructing the model, the influence of each component
of the galaxy in the CMD is presented. This can be
seen in Figure~\ref{figmodel}, the top panels show the (61,-15)$^\circ$
field from the model being split into its various Galactic components.
For $g_\circ$ and $r_\circ$ CMDs, this plot illustrates where the various
components of the Galaxy lie on the CMD.  The lower panels are from
the model field (75,+15)$^\circ$, and show how a ($V_\circ$,$i_\circ$) CMD can be
similarly deconstructed.  Analysing Figure~\ref{figmodel}, shows
that each component of the Milky Way occupies different regions of the
CMD. In the following subsections the equations defining the separate
Galactic structures as used in the synthetic galaxy model\footnote{
  Each of the following equations is cited and presented as per the
  referenced papers. A slight modification has been made to the Thick
  Disk equation to account for a typographical error in the
  paper.} are presented.
\subsubsection{Thin Disk density profile}
The youngest disc\footnote{\citet{1986A&A...157...71R}}:
$R$ and $z$ are the Galactocentric cylindrical coordinates (pc), with $c$
as the axis ratio for the ellipsoidal components. A is a constant
defined by Table 1 of \citet{1986A&A...157...71R}.
\begin{equation}
a^2 = R^2 + \frac{z^2}{c^2} \notag
\end{equation}
\begin{equation}
\rho(R,z) = A\biggl\{{exp{\bigl(-\bigl(\frac{a}{K_{\text{\tiny{+}}}}\bigr)^2}\bigr)} - {exp{\bigl(-\bigl(\frac{a}{K_{\text{\tiny{-}}}}\bigr)^2}\bigr)}\biggr\}
\end{equation}
with $K_+ = 5000pc$, $K_- = 3000pc$ and $c = 0.014$

Old discs:
\begin{equation}
  \begin{split}
\rho(R,z) = A\Biggl\{exp{\biggl(-\biggl(0.5^2 +
  \bigl(\frac{a^2}{K_{\text{\tiny{+}}}^2}\bigr)\biggr)^\frac{1}{2}\biggr)}\\
 - {exp{\biggl(-\biggl(0.5^2 + \bigl(\frac{a^2}{K_{\text{\tiny{-}}}^2}\bigr)\biggr)^\frac{1}{2}}\biggr)}\Biggr\}
  \end{split}
\end{equation}
with $K_+ = 2226pc$ and $K_- = 494.4pc$ The Thin Disk is quite tightly constrained in the $g$,$r$ CMD but is
quite broad in the $V$,$i$ CMD, it turns redward at $\sim$19$^{th}$ magnitude in
the $g$,$r$ CMD, while for the $V$,$i$ CMD it is not as well defined and blurs
down to $\sim$20$^{th}$ magnitude. 

\subsubsection{Thick Disk density profile}
\begin{equation}
  \rho(R,z) \propto
  \begin{cases}
    exp{\bigl(-\frac{R - R_{\odot}}{h_{R}}\bigr)} \bigl(1-\frac{1/h_{z}}{x_{l}(2 + x_{l}/h_{z})} z^2\bigr)& \text{if $z$ $\leqslant$ $x_{l}$}\\
    exp{\bigl(-\frac{R - R_{\odot}}{h_{R}}\bigr)} exp{(-\frac{z}{h_{z}})}& \text{if $z$ $>$ $x_{l}$}
  \end{cases}
\end{equation}
The Thick Disk\footnote{\citet{2001A&A...373..886R}} equations have
three parameters defining the density along the $z$ axis: $h_z$, the scale height, $\rho_{\circ}$, the local density and $x_l$ the
distance above the plane where the density law becomes
exponential. This third parameter is fixed by continuity of
$\rho$($z$) and its derivative.  It varies with the choice of scale
height and local density following the potential. These variations are
listed below\footnote{\citet{1996A&A...305..125R}},
\begin{equation}
  \begin{split}
    x_{l} = 1358.6 - 1.35nh + 2.335^{-4}nh^2 \\
    - nr(8.1775^{-1}+5.817E(-3)nh)
  \end{split}
\end{equation}
where the scale height density is
\begin{equation}
     nh = \frac{h_{z}}{1 pc}
 \end{equation}
and the local density is
\begin{equation}
     nr = \frac{\rho_{\circ}}{1.22E(-3){stars.pc^{-3}}}
 \end{equation}
From Table 1 of \citet{1996A&A...305..125R} the scale height is 
$h_{z}$ = 760 $\pm$ 50 pc, the local density $\rho_{\circ}$ = 5.6
$\pm$ 1\% and the scale length $h_{R}$ = 2.8 kpc. The Thick Disk in the $g$,$r$ CMD is well constrained at the ends of the magnitude range
(keeping in mind that these figures do not take into account any
incompleteness) and has a redward trend around 21$^{st}$ magnitude.  The
Thick Disk in $V$,$i$ is much more diffuse and the redward trend extends to
$\sim$22$^{nd}$ magnitude.

\subsubsection{Halo density profile}
\begin{equation}
\rho(R,z) = \rho_{\circ} (R^2 + \frac{z^2}{\varepsilon^2})^\frac{n}{2}
\end{equation}
In the Halo \footnote{\citet{2000A&A...359..103R}} or spheroidal
equations, $\rho_{\circ}$ is the local density, $n$ is the power law
index and $\varepsilon$ is the flattening. The best fit values for
these parameters ~\citep{2000A&A...359..103R} are $\rho_{\circ}$ =
1.64E(-4) stars.pc$^{-3}$, $n$ = 2.44 and $\varepsilon$ = 0.76. The Halo in the $g$,$r$ CMD seems to have a very faint
trend redwards at $\sim$24$^{th}$ magnitude, however in the $V$,$i$ CMD there is a distinct
turn redward in the Halo around 22$^{nd}$ magnitude. 

These figures illustrate why the Galactic component of the CMD does not introduce any strong main
sequences into the CMD.  Instead, they are smooth with overdensities
at the extremes of the magnitude range, with each component
introducing redward trends at different places in the CMD.  Understanding the CMDs in this
way allows for greater understanding of the data when interpreting the
observations.

CMDs of the data were produced using the matched catalogues from the
data-reduction pipeline, shown as Hess plots\footnote{A Hess plot is a
  pixelated Colour-Magnitude diagram where the grayscale denotes the
  square root of the Colour-Magnitude diagram number density} in
Figures~\ref{fig61pm15} to~\ref{fig150pm15}. The Galactic
coordinate notation from here on will be written in the form ({\it l},
$\pm${\it b})$^\circ$. In Figures~\ref{fig61pm15} to~\ref{fig150pm15}, the
top-left panel is the Northern field, the top-right panel is the Southern
field, each containing the actual data from the observing run.  The
lower panels are solely synthetic CMD's and are again North on the
lower-left and South on the lower-right.

All previous detections of the MRi have shown it to be beyond the Disk
of the galaxy.  This means that the MRi main sequence should be in the
area of CMD typically containing the Halo and Thick Disk stars. Since
the density of stars here is intrinsically low, the MRi main sequence
should contrast well against these weaker components. However, if the
stream approaches close to the edge of the Thick Disk, there is the
possibility of confusion between the two.  All of the possible
detections of the stream that result from this work are in this region
between the Thick Disk and the Halo and so there remains some
ambiguity over how to separate these two structures from the stream.
Each of the detections represents large excursions from the
synthetic galaxy model, typically $\sim$1 magnitude or greater, corresponding to a
distance shift of $\sim$6 kpc according to our estimation.  Comparisons with the
model in this manner is unsuited towards minor deviations, but given
the distances involved in a $\sim$1 magnitude difference this suggests
the features represent more than just problems with the synthetic
galaxy model.  If this were the case then the model is indeed very
inadequate in these regions.  Since we do not expect this to be the
case we proceed with this method. Ultimately, kinematic surveys of these regions will be needed to
resolve some of the remaining uncertainties but in the meantime we have used careful
comparison with the models of \cite{2003A&A...409..523R} coupled with
completeness and magnitude-error estimates to draw out any extra
features from the Colour-Magnitude diagrams.

\subsubsection{Magnitude Completeness}\label{completeness}
The magnitude completeness of the data was estimated using the overlap regions
of the observations.  Each field consists of several overlapping
subfields (see Table~\ref{ObsTable}). Stars in the overlap regions
will appear in more than one subfield.  So by dividing the number of
stars that can be matched across subfields by the total number of
stars in the overlap region (for a given magnitude bin), the
completeness curve can be formed. Fitting the curve with the function,
\begin{equation}
  CF = \frac{1}{1 + e^{(m - m_c)/ \lambda}}\\
\end{equation}
where m is the magnitude of the star, m$_c$ is the magnitude at 50\%
  completeness and $\lambda$ is the width of the rollover from 100\%
  completeness to 0\% completeness. By applying the completeness curve
  of the data to the model makes for a more suitable comparison.
  Our method for determining completeness does not account for faint
  stars which lie close to bright stars and thus are excluded from our
  completeness estimate. The values used to model each field can be
  found in Table~\ref{CompTable}.
  
\begin{table}
\centering
\caption{Parameters used to model the completeness of each field,
  ordered in ascending Galactic longitude ({\it l}). See
  $\S$\ref{completeness} for a description of each column. }
\begin{tabular}{||r|c|c|c||} \hline \hline
Fields \it{(l,b)$^\circ$} & m$_c$  ($g_\circ$
or $V_\circ$)  & m$_c$ ($r_\circ$
or $i_\circ$)  & $\lambda$ \\ \hline
 (61,$+$15)$^\circ$   & 24.2 & 23.2 & 0.4 \\ \hline
 (61,$-$15)$^\circ$   & 24.5 & 23.5 & 0.4 \\ \hline
 (75,$+$15)$^\circ$   & 23.6 & 22.5 & 0.5 \\ \hline
 (75,$-$15)$^\circ$   & 22.8 & 21.8 & 0.5 \\ \hline
 (90,$+$10)$^\circ$   & 23.2 & 22.2 & 0.3 \\ \hline
 (90,$-$10)$^\circ$   & 23.2 & 22.2 & 0.3 \\ \hline
(118,$+$16)$^\circ$   & 23.4 & 22.4 & 0.4 \\ \hline
(123,$-$19)$^\circ$   & 24.0 & 23.0 & 0.5 \\ \hline
(150,$+$15)$^\circ$   & 23.0 & 21.5 & 0.3 \\ \hline
(150,$-$15)$^\circ$   & 23.2 & 22.0 & 0.3 \\ \hline
\hline
\end{tabular}
\label{CompTable}
\end{table}

With regards to the real data, the uncertainty in the magnitude
increases with increasing magnitude.  This
explains why the structures in the real data loses coherency at the
faintest extremes of the Hess plots.  By plotting the magnitude
against the error of the magnitude it can be fitted with the following
function. 
\begin{equation}
    f(x) = A + Be^x + Cx^2
\end{equation}
where {\it f(x)} is the error in the magnitude and {\it x} is the
magnitude\footnote{Parameters for these equations will be supplied
  upon request (bconn@physics.usyd.edu.au)}.

This allows a similar effect to be introduced into the model figures
when making quantitative measurements. The completeness and magnitude
error functions have not been applied when making initial comparisons
between the model and the data in a qualitative sense. Comparing
obvious structural differences with the model and finding the distance
to any new structure is a qualitative approach and does not require the model to be corrected in this
manner. Then, when attempting to find the signal-to-noise of our detections
quantitatively, we have applied both corrections to the model adjusting it to match the data.

\subsubsection{Estimating the Distance}\label{distance}
Using the method employed by \citet{2003MNRAS.340L..21I}, we too have
used the colour-transformation which converted the ridge-line of the
SDSS S223+20 field [~\citet{2002ApJ...569..245N}, see Figure 12.] to a main sequence type overlay.  By determining the
offset of any new structure from the base position of this ridge-line
we can estimate the distance. To do this we need to convert the
SDSS ($g'$,$r'$) system to the Vega-normalized ($g$,$r$) and
($V$,$i$), this can be done by comparing overlapping INT and SDSS
fields (SDSS fields taken from Early Data Release,
\citet{2002AJ....123..485S}).  In particular, these conversions use a
comparison field near the
Galactic South Pole.  The relevant colour conversions are:

for $g$ and $r$, 
\begin{equation}
    (g-r) = 0.21 + 0.86(g'-r')
\end{equation}
\begin{equation}
  g = g' + 0.15 - 0.16(g-r)
\end{equation}

for $V$ and $i$,
\begin{equation}
    V = g -0.03 - 0.42(g-r)
\end{equation}
\begin{equation}
  (g-i) = 0.09 + 1.51(g-r)
\end{equation}

The ($V$,$i$) conversion is taken from
\citet{1991ApJ...380..362W}\footnote{These conversions may have
  significant systematic errors and as such any distance estimate
  derived using this method should merely be taken as indicative in particular, the conversion from ($g'$,$r'$) to ($V$,$i$).}
The zero offset Heliocentric distance estimate is assumed to be 11 kpc
\citep{2002ApJ...569..245N} and any deviation in magnitude required to
align this main sequence style overlay is assumed to be solely due to
distance variations. The Heliocentric distance is then calculated
using Eqn.\ref{helio} and the Galactocentric distance is found using
simple trigonometry assuming the Sun is located 8.0 kpc from the
Galactic centre.

\begin{equation}\label{helio}
  R_{HC} = 11.0\bigl(10^\frac{offset}{5.0}\bigr)
\end{equation}

Determining a value for the error
associated with such a measurement is dependent on several factors.
Most predominant of these is whether the fields have been correctly
calibrated with regard to their photometry taking into account the
dust extinction present within the fields
\citep{1998ApJ...500..525S}. Given however that the accuracy of the
dust maps is $\sim$16\%, this will dominate over the few percent
calibration error in the photometric zero-points as derived by the CASU
pipeline ~\citep{2001NewAR..45..105I}.  Having understood the errors
involved in both the determination of the photometry, extinction correction and the
main-sequence style overlay, manually placing this overlay at the two
extremes of an acceptable fit provides a range of distances over which
this structure resides. The final value then represents the average of
this estimate rounded to the nearest kiloparsec; the error in the fit is typically less than 1 kpc and
thus is dominated by the $\sim$10\% error which naturally resides in the original
distance calculation. Given the large errors involved these distances
can only be considered indicative of the true distance.

\subsection{Individual Fields}\label{Individual Fields}

\subsubsection{Fields \bf$(61,\pm15)^\circ$}\label{61pm15des}

\begin{figure}\centering
\includegraphics[width=83mm]{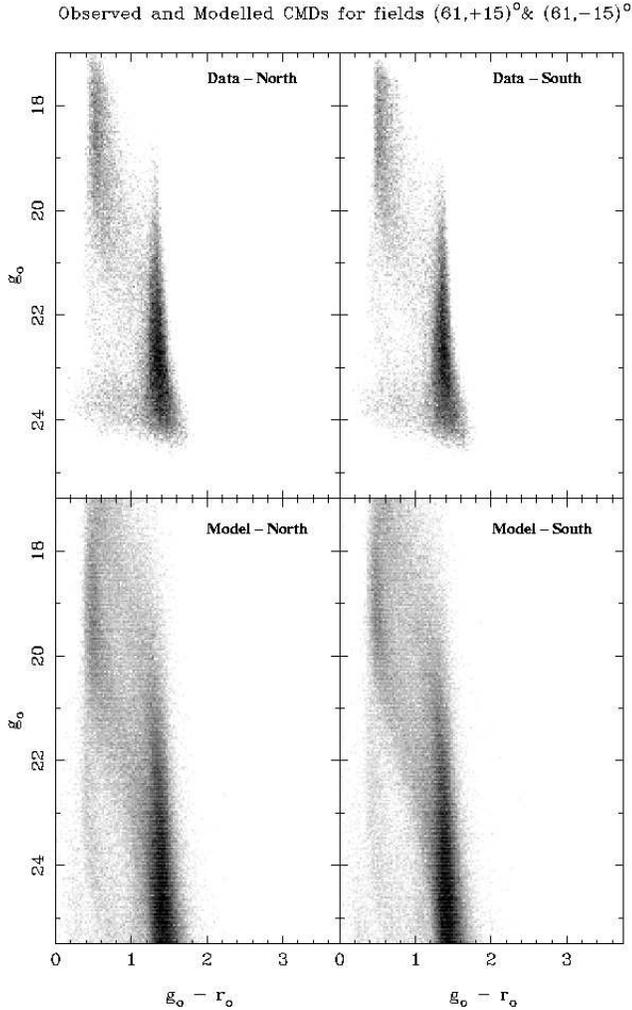}
\caption[]{Hess plots  of ({\it l},{\it b}) = (61,+15)$^\circ$ and
  (61,-15)$^\circ$ with the accompanying model CMDs. The top panels
  show the data obtained by the INT/WFC, visually it can be seen that the
  limiting magnitudes for the data here is $g_\circ$$\sim$23.8.  The
  model CMDs presented in the lower panels have not been corrected for
  completeness and as such continue below the limiting magnitudes of
  the data.
\label{fig61pm15}}
\end{figure}

Presented in the top panels of Figure~\ref{fig61pm15}, the data is
shown as a Hess plot.  The lower panels were generated by the synthetic galactic model of
\citet{2003A&A...409..523R} to serve as comparison figures for the
data. Note that the model fields have not been completeness corrected
for these comparisons. When comparing with Figure~\ref{figmodel},
we can see each of the expected Galactic components present in the
data. Thus the observed field matches the synthetic field to a high degree
supporting our assumption that the synthetic fields will act as good
comparisons to the data.  In these fields, the Halo component is
seemingly showing a redward trend brighter than the model, however the strength of
this feature was deemed not sufficiently significant to infer a detection.
$\S$\ref{penacomp2} re-examines this field in light of the model
produced by \cite{Pena2004}.

\subsubsection{Fields \bf$(75,\pm 15)^\circ$}\label{75pm15des}

\begin{figure}\centering
\includegraphics[width=83mm]{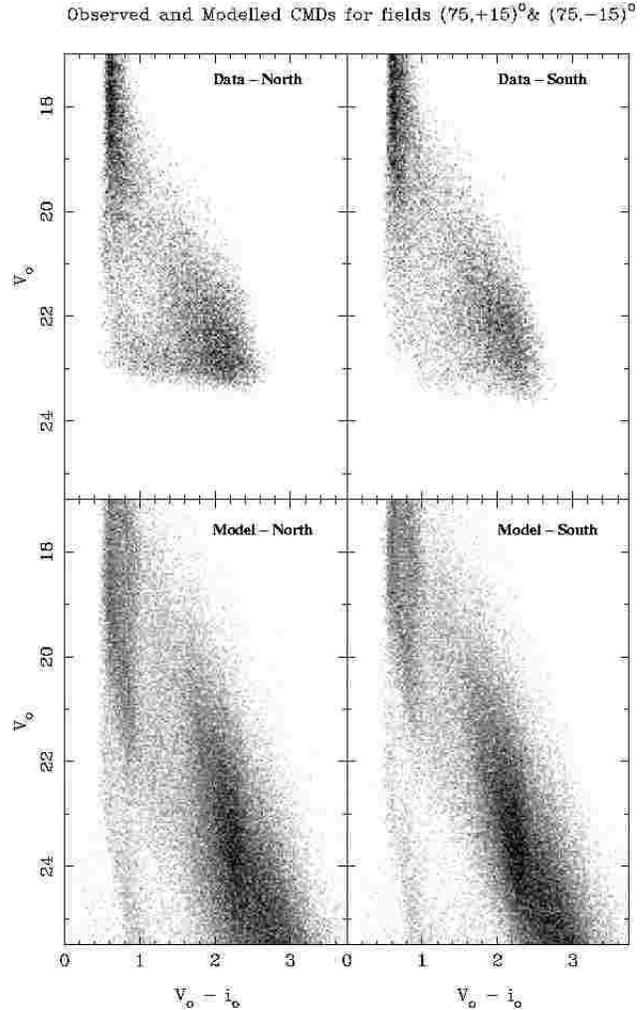}
\caption[]{Hess plots  of  ({\it l},{\it b}) = (75,+15)$^\circ$ and
  (75,-15)$^\circ$. The limiting magnitude in this figure is $V_\circ
  \sim$23.6 for the Northern field and $V_\circ \sim$22.8 for the Southern
  field. The figure is otherwise the same as Figure~\ref{fig61pm15}.
\label{fig75pm15}}
\end{figure}

Figure~\ref{fig75pm15} shows the data and the model for the fields
(75,$\pm$15)$^\circ$.  These fields were observed using the $V$ and $i$ filters and this has
the effect of changing the layout of the CMD.  Studying
Figure~\ref{figmodel}, each of the Galactic components in our data can
again be easily identified .  The field (75,+15)$^\circ$ also seems to
have a Halo component which turns redward brighter than the model. Similarly
with the fields (61,$\pm$15)$^\circ$ this feature is not significant enough to
convincingly label a detection.  These fields too are revisited
in $\S$\ref{penacomp2} when comparing with the results of \cite{Pena2004}.

\subsubsection{Fields \bf$(90,\pm10)^\circ$}\label{90pm10des}

\begin{figure}\centering
\includegraphics[width=82mm]{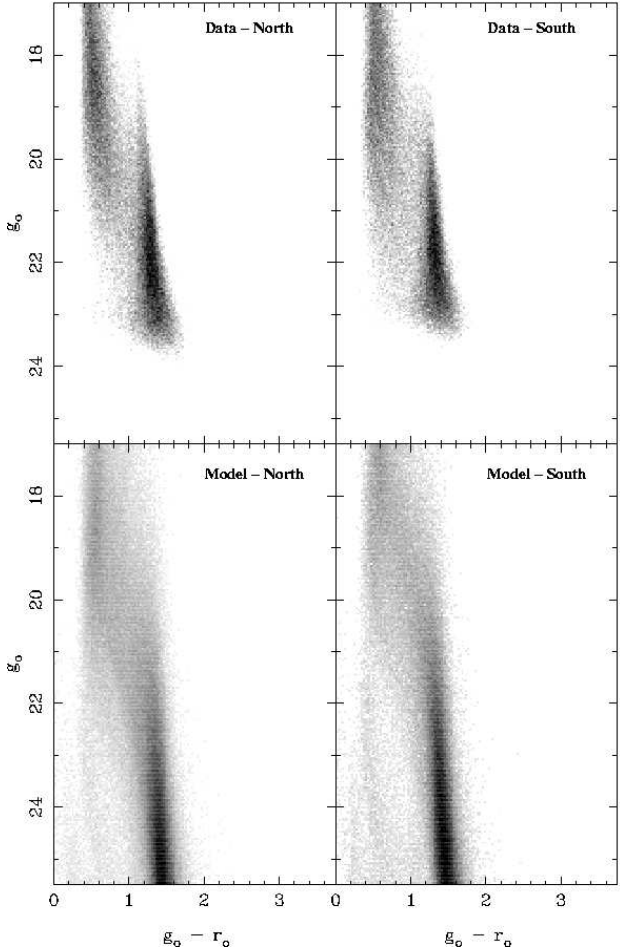}
\caption[]{Hess plots  of ({\it l},{\it b}) = (90,+10)$^\circ$ and (90,-10)$^\circ$.
  As for Figure~\ref{fig61pm15}. 
\label{fig90pm10}}
\end{figure}

Figure~\ref{fig90pm10} shows the data and the model for the fields
(90,$\pm$10)$^\circ$.  As with Figure~\ref{fig61pm15} the layout is the
same and also was observed in $g_\circ$ and $r_\circ$, possessing similar
Galactic CMD structure.  Again, visual inspection reveals no extra structure in this CMD and hence no MRi
signature.

\subsubsection{Fields {\bf$(118,+16)^\circ$} \& {\bf$(123,-19)^\circ$}}\label{118p16des}
\begin{figure}\centering
\includegraphics[width=83mm]{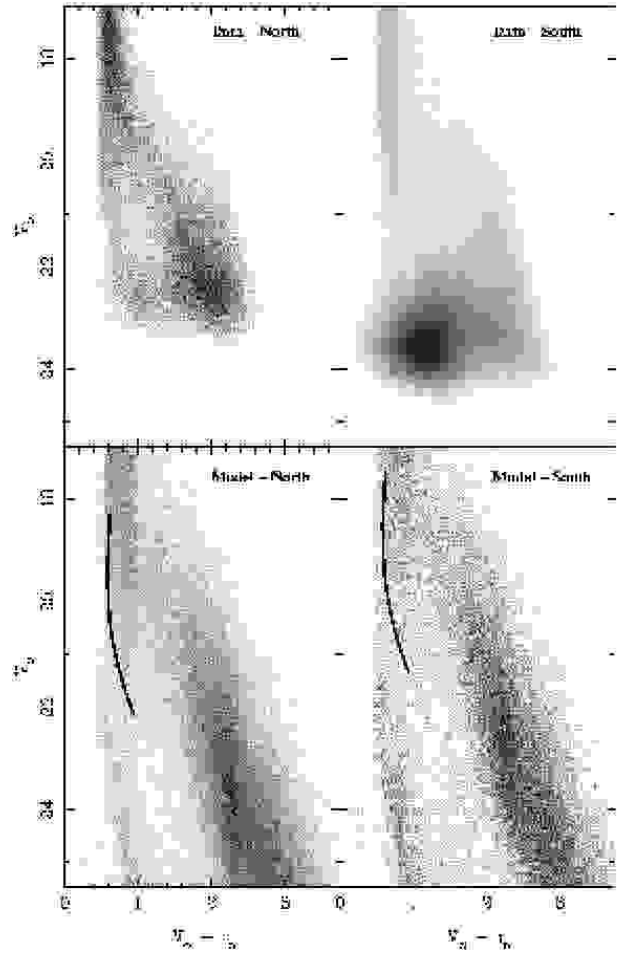}
\caption[]{ As for Figure~\ref{fig61pm15}, Hess plots of ({\it l},{\it
  b}) = (118,+16)$^\circ$ and (123,-19)$^\circ$
  ~\citep[from][]{2003MNRAS.340L..21I}.  The  Heliocentric distances
  (R$_{HC}$) of   the main sequence overlays are $\sim$12 kpc for the
  Northern field from an offset of +0.2 and $\sim$7.6 kpc for
  the Southern field from an offset of -0.8, with
  Galactocentric distances (R$_{GC}$) being $\sim$17 kpc and
  $\sim$14 kpc. The dense region located at
  ($V_\circ\sim23.5,V_\circ-i_\circ\sim1.2$) is M31 in the background of this field.
\label{fig118p16}}
\end{figure}

Figure~\ref{fig118p16} shows the data and the model for the fields
(118,+16)$^\circ$ and (123,-19)$^\circ$. The (123,-19)$^\circ$ field is from the
M31 survey \citep{2001Natur.412...49I,2002AJ....124.1452F} with the
confirmation of the MRi signature presented in
~\citet{2003MNRAS.340L..21I}. The (118,+16)$^\circ$ field has a Halo
component in which the Main Sequence (MS) turns redward brighter than that of the
model field. As from $\S$\ref{75pm15des} the bulk Halo MS is expected to
turn redward in this field at $V_\circ\sim22.5$. The redward trend detected occurs at
$V_\circ\sim20.0-21.0$ and reaches a value of $V_\circ-i_\circ\sim1.0$ at
$V_\circ\sim23.0$.  Comparing this to the expected location of the Halo
population in the model reveals that the (118,+16)$^\circ$ field is inside this
population and thus is a potential detection of the MRi.  Fitting a
main sequence to the feature and overlaying it on the model, as can be
seen in the lower left plot of Figure~\ref{fig118p16}, an
offset of +0.2 magnitudes or R$_{HC}$ $\sim$12 kpc is calculated which
corresponds to R$_{GC}$ $\sim$17 kpc.  This is further out than is predicted by the model shown in
Figure~\ref{figaitoff}.  The field (118,+16)$^\circ$ is intended to
be a comparison field on the symmetrically opposite side of the
Galaxy.  Both these fields have been observed in $V$ and $i$ and thus have
the Galactic features as discussed for Figure~\ref{fig75pm15}.  The
MRi signature in (123,-19)$^\circ$ stands out clearly against the known
Galactic components.  The magnitude offset is -0.8 with distances to this feature being R$_{HC}$
$\sim$7.6 kpc and R$_{GC}$ $\sim$14 kpc.  Note that
the circular overdensity at ${\rm V_\circ\sim23.5},{\rm V_\circ-i_\circ\sim1.2}$
is M31 in the background of this field.
\subsubsection{Fields \bf$(150,\pm15)^\circ$}\label{150pm15des}

\begin{figure}\centering 
\includegraphics[width=83mm]{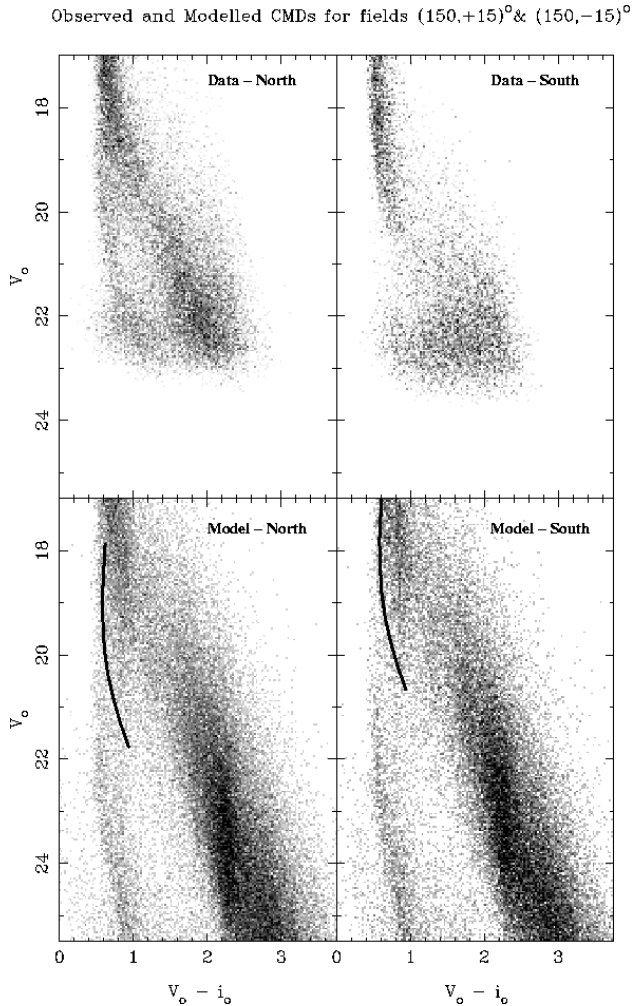}
\caption[]{Hess  plots of  ({\it l},{\it b}) = (150,+15)$^\circ$ and
  (150,-15)$^\circ$. As for Figure~\ref{fig61pm15}. The main sequences of
  the potential detections are shown as overlays on the synthetic
  CMDs however note that the detection in the Southern field is very
  tentative. The Heliocentric distance to the Northern field detection is
  $\sim$9 kpc and $\sim$6 kpc for the Southern
  field calculated from offsets of -0.4 and -1.5 respectively.  Galactocentrically, the distances are, $\sim$17 kpc
  and $\sim$13 kpc.
\label{fig150pm15}}
\end{figure}

Figure~\ref{fig150pm15} shows the data and the model for the fields
(150,$\pm$15)$^\circ$. These fields have been observed in $V$ and $i$ and thus
have the Galactic features as discussed for Figure~\ref{fig75pm15}.
Using the technique described in $\S$\ref{118p16des} of examining
when the bulk Halo Main Sequence (MS) turns redward, in the Northern field this
occurs at $V_\circ\sim22.5$. The data MS turns redward at $V_\circ\sim21.0$ and
reaches a value of $V_\circ-i_\circ\sim1.2$ at $V_\circ\sim23.0$.  Fitting a main
sequence to this feature and overlaying it on the model using an
offset of -0.4 magnitudes, which can be
seen in the lower left plot of Figure~\ref{fig150pm15}, we calculate
R$_{HC}$ $\sim$9 kpc, this corresponds to R$_{GC}$
$\sim$17 kpc which exceeds the model prediction in
Figure~\ref{figaitoff} by $\sim$7 kpc. Comparing with the detection in
field Mono-N (149,+20)$^\circ$, ~\citet{2003MNRAS.340L..21I} shows
similar findings.  They also report an offset of -0.4 magnitudes
corresponding to a Heliocentric distance of $\sim$9 kpc and a
Galactocentric distance of $\sim$16 kpc.  In all likelihood this
represents two detections of the same stream in different passbands
separated by a few degrees on the sky, given this, we have interpreted
it as a detection of the MRi.  The (150,-15)$^\circ$ field only vaguely resembles
its synthetic field counterpart, however the strong vertical feature
at $V_\circ-i_\circ\sim0.5$ seems to form a main sequence that is quite close
to the edge of the Thick Disk in the current Galactic model.
Determining whether this is in fact the edge of the Thick Disk is
difficult considering that the Halo component is poorly defined and
thus not readily available for comparison.  Fitting a main sequence to
this feature using an offset of -1.5,
R$_{HC}$ $\sim$6 kpc and R$_{GC}$ $\sim$13 kpc. 

\subsection{A re-examination of field \bf$(123,-19)^\circ$}\label{123m19}
\begin{figure}\centering
\includegraphics[width=83mm]{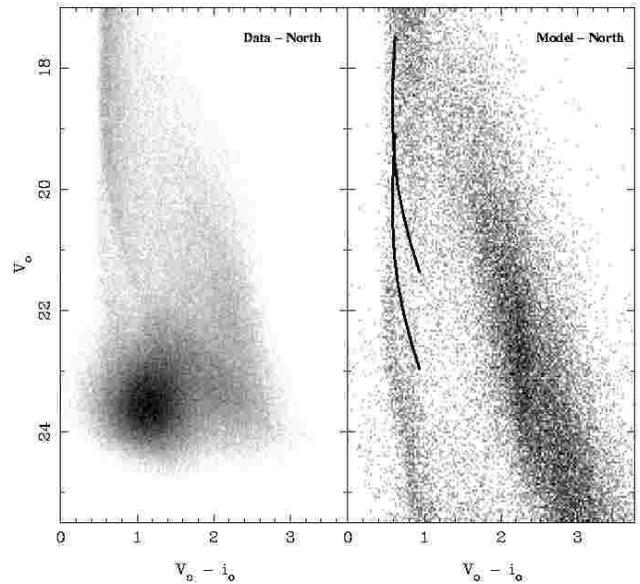}
\caption[]{Hess  plots of  ({\it l},{\it b}) = (123,-19)$^\circ$ and its
  counterpart synthetic CMD. The main sequences of
  the potential detections are shown as overlays on the synthetic
  CMDs. The Heliocentric distances to the higher main sequence is (as
  before) $\sim$8 kpc and the lower main sequence
  $\sim$16 kpc.  The Galactocentric distances being $\sim$14
  kpc and $\sim$21 kpc.  Clearly, the lower main sequence
  deviates from the model of the Halo  and is separate from known
  Galactic components.
\label{fig123m19}}
\end{figure}

Detecting the stream by noting deviations from the synthetic CMDs as in the (118,+16)$^\circ$ and
(150,$\pm$15)$^\circ$ fields, rather than the obviously additional feature as seen in
Figure~\ref{fig118p16} with regards to the (123,-19)$^\circ$ field,
means applying this technique to the (123,-19)$^\circ$ field may yield a new detection. 
The redward trend of the Halo component begins at $V_\circ\sim$ 22-22.5 in the
model while in the data this occurs at $V_\circ\sim$ 21-21.5, suggesting this feature is not of Galactic
origins.  Fitting a main sequence overlay with an offset of +0.8 to this, as can be seen in
Figure~\ref{fig123m19}, distance estimates of  R$_{HC}$
$\sim$16 kpc and R$_{GC}$$\sim$21 kpc are obtained
for this new structure. 

If however, this is to believed, then there are two structures present in this
field.  Since this field is in the vicinity of the newly discovered
TriAnd structure perhaps this is a detection of it in the background.
The current distance estimate to the TriAnd structure is
R$_{HC}$$\sim$15-30 kpc, in the region {\it l} $\sim$100$^\circ$ to $\sim$170$^\circ$
and {\it b} $\sim$-15$^\circ$ to $\sim$-60$^\circ$.  This places both
detections in this field on the edge
of this structure. Although association between the MRi and TriAnd is still speculative,
TriAnd may represent a wrapped tidal arm of the MRi. Pursuing
this thought, a wrapped tidal arm would imply that the MRi is not a
recent accretion event but rather the relic of a much older
accretion. The model of \cite{Pena2004}, also supports this view of wrapped
tidal arms and this field in particular fits with their
model. $\S$\ref{penacomp1} discusses this in further detail.  Clearly
though, this is a tentative detection as the correspondence between the
model and the data, in this region, is poor.  This detection may
also simply reveal a deficiency in the synthetic galaxy model.

\subsection{Comparison: Data minus Model}\label{subtraction}

\begin{figure}
\centerline{ \psfig{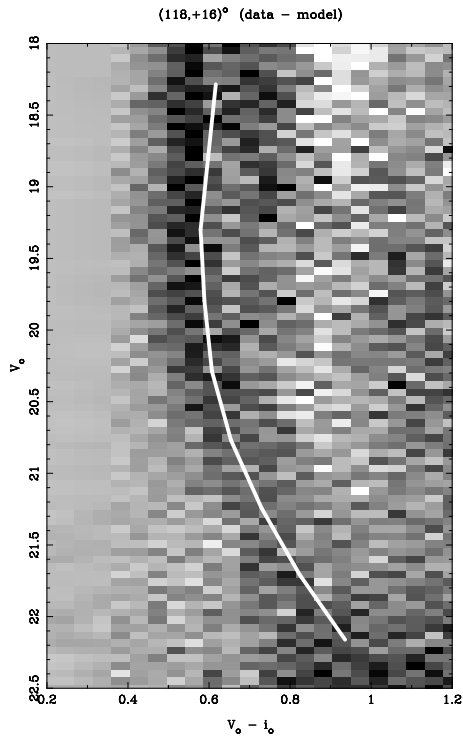}}
\caption[]{Close up of field (118,+16)$^\circ$ (Figure~\ref{fig118p16}) after
  subtracting the completeness and magnitude error corrected number CMD
  of the model in the same region. The main sequence overlay in this
  figure is the same as the overlay presented in Figure~\ref{fig118p16},
  with the signal-to-noise ratio estimate of this feature to be $\sim$20.
\label{fig118p16zoom}}
\end{figure}

\begin{figure}
\centerline{ \psfig{figure=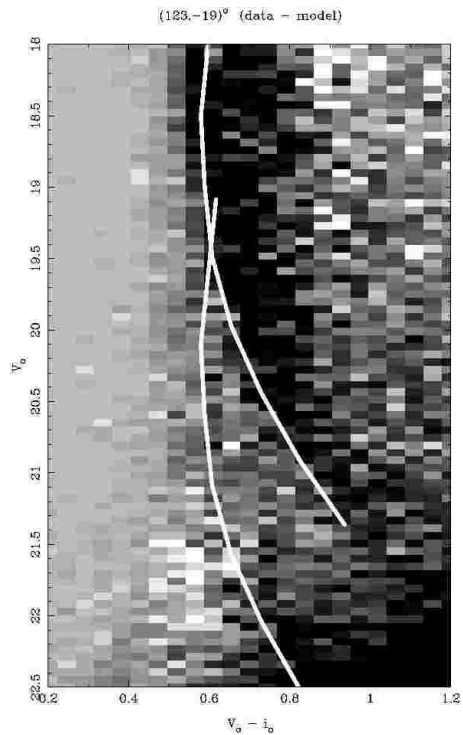,angle=0,width=6cm}}
\caption[]{Close up of field (123,-19)$^\circ$
  (Figure~\ref{fig123m19}), in the same manner as
  Figure~\ref{fig118p16zoom}. The main sequence overlay in this figure
  is the same as overlay presented in Figure~\ref{fig118p16}. The
  first, being the main sequence presented in
  \citet{2003MNRAS.340L..21I} determined to have a signal-to-noise
  ratio of $\sim$20, in accordance with \citet{2003MNRAS.340L..21I}.
  The second, our ``by-eye'' detection of a more distant stream, is
  much more tentative and a determination of a signal-to-noise ratio
  proved far more difficult.  A discussion on these overlays is
  presented in $\S$\ref{123m19}.
\label{fig123m19zoom}}
\end{figure}

\begin{figure}
\centerline{ \psfig{figure=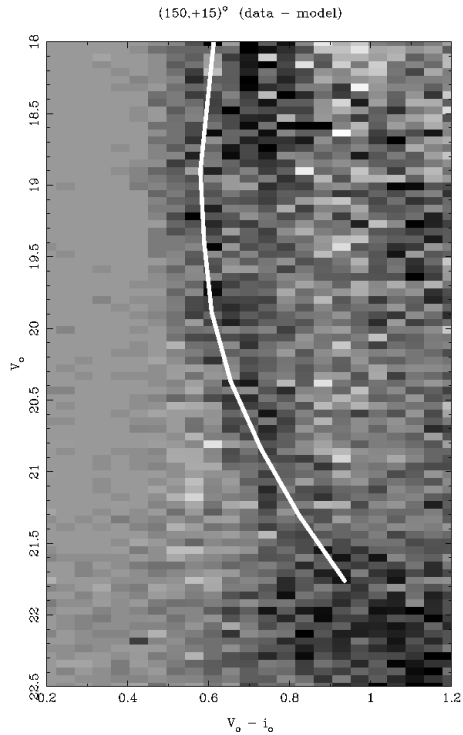,angle=0,width=6cm}}
\caption[]{Close up of field (150,+15)$^\circ$
  (Figure~\ref{fig150pm15}), in the same manner as
  Figure~\ref{fig118p16zoom}. The main sequence overlay in this
  figure is the same as overlay presented in Figure~\ref{fig150pm15},
  the signal-to-noise ratio of this feature is around $\sim$12.
\label{fig150p15zoom}}
\end{figure}

Applying the completeness corrections and the appropriate magnitude
error estimates to the model brings it as close to the real data as
possible. By subtracting the corrected model from the data this minimises the residuals between the
two. For those fields in which we found no ``by-eye''
detection, namely, (61,$\pm$15)$^\circ$, (75,$\pm$15)$^\circ$ and (90,$\pm$10)$^\circ$, the residuals
formed no coherent structures and have not been presented here. The
field (150,-15)$^\circ$ is also not shown, as the differences between the real
data and the model proved too great to draw any logical conclusions. The
remaining fields (118,+16)$^\circ$, (123,-19)$^\circ$ and (150,+15)$^\circ$ are shown in
Figures~\ref{fig118p16zoom},~\ref{fig123m19zoom} and \ref{fig150p15zoom}. The
same main sequence lines as drawn in the previous Hess plots have also
been overlaid on these CMDs.  The features as mentioned previously
are present in these figures also, see $\S$\ref{Individual Fields}.

This procedure allowed us to quantitatively determine an estimate of the significance
 of these features, found by dividing the number of stars in the
 feature by the Poisson noise due to the stars in the region.  The
 feature in Figure~\ref{fig118p16zoom}, from (118,+16)$^\circ$, has a signal-to-noise ratio of
 $\sim$20. For field (123,-19)$^\circ$ (Figure~\ref{fig123m19zoom}),
 the estimated signal-to-noise value is $\sim$20, which is in accordance with
 ~\citet{2003MNRAS.340L..21I}.  Finally for (150,+15)$^\circ$
 (Figure~\ref{fig150p15zoom}), the signal-to-noise was found to be
 $\sim$12. While we may not know the origins of these features, these results
confirm that they are significant contributors to the CMDs in these regions.

\section{Discussion}\label{conclusions}

\begin{figure}
\centerline{ \psfig{figure=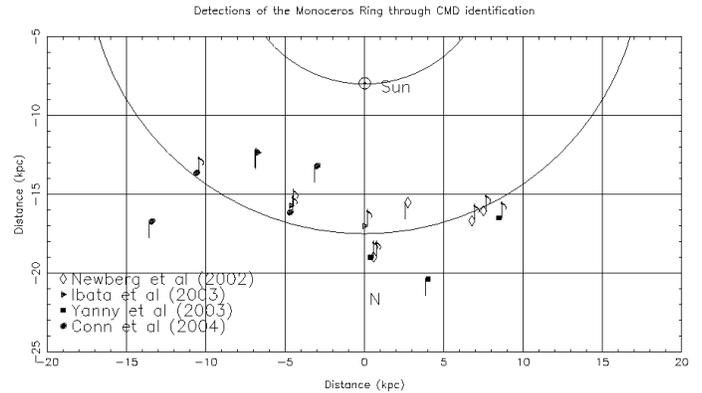,angle=270,width=9cm}}
\caption[]{Galactic View of the Monoceros Ring through CMD
  identification. Shown in a top down Galactic view, the Sun is at
  (0,-8).  The quavers (notes with tails up) are detections above the plane of the galaxy ({\it b} $>$
  0$^\circ$) and the crochets (notes with tails down) are detections below
  the plane of the Galaxy ({\it b} $<$ 0$^\circ$). The symbols denote the
  different observations. The inner circle marks out the solar circle
  around the Galaxy at 8 kpc and the outer circle has a radius of 17.5
  kpc.  The line joining the `N' through the Sun corresponds to the
  axis of the node of the warp according to ~\citet{2003A&A...409..523R}. In
  this projection, Galactic longitude {\it l} is measured
  anti-clockwise from the line joining the Sun to the Galactic centre.
\label{figMRidetect}}
\end{figure}
\subsection{Summary of Results}

Using the Isaac Newton Telescope Wide Field Camera in La Palma, Canary
Islands we have undertaken a survey of the Monoceros Ring (MRi) in the
region of Galactic longitudes {\it l} = (61 - 150)$^\circ$ in symmetric
pairs above and below the plane.  Table~\ref{ObsTable} shows the
fields observed. The MRi can be seen in many fields across the sky
(Figure ~\ref{figMRidetect}) and also in 2MASS distribution of M-giant stars
\citep{2003ApJ...594L.115R}.  The origin of this structure is still
poorly understood and a systematic survey around the Galactic plane is
required to understand the size and extent of this structure.

Surveying the region of sky {\it l}= (61 - 150)$^\circ$ and through a
re-examination of field (123,-19)$^\circ$, four tentative detections of the
MRi have been found. Fitting main sequences to those CMDs which showed a deviation from the
synthetic galaxy model has yielded distance
estimates in those fields. The distance estimates both
Heliocentric and Galactocentric can be seen in Table~\ref{DistTable}.

With detections in each of these cases being on both sides of the
Galactic plane it suggests the stream is either very broad on the
plane of the sky or represents more than one arm of the
tidal debris wrapping around on themselves.  Also, the detection at (150,+15)$^\circ$
is on the opposite side of the Galactic plane to the newest structure
in the Milky Way, namely the TriAnd feature of \citet{2004ApJ...615..732R}.

Are they both related to the MRi structure? The kinematics of most of
the MRi detections are unknown, however with more
detections of the MRi on both sides of the plane, the MRi structure is
possibly much older than previously assumed. If indeed it is an
accreted dwarf galaxy, then it has seemingly made several orbits
around the galaxy.

\subsection{Re-examination of the observations based on the model of \citet{Pena2004}}
The hypothesis of multiply wrapped tidal arms is investigated by
\citet{Pena2004} who have created an detailed model of the
Monoceros tidal stream using all the data currently available.
By testing thousands of models of tidal streams, in both prograde and
retrograde directions around the Milky Way and correlating them
against the current observations they refined the parameters for the
tidal stream. The solution they favour is one where the accreting dwarf galaxy has
made two orbits of the galaxy in the prograde direction leaving two concentric tidal
streams. 
\subsubsection{Detections: Fields (118,+16)$^\circ$,(123,-19)$^\circ$ \& (150,$\pm$15)$^\circ$}\label{penacomp1}
Interestingly, Figure 7 of \cite{Pena2004} shows the location of the stream in Galactic coordinates and the
corresponding Heliocentric distances to the stream around the Galactic
equator, allowing for a direct comparison with our data. At
(118,+16)$^\circ$ their model shows the stream around $\sim$17 kpc
away and at the location (123,-19)$^\circ$ shows a region where the two streams
cross, the inner ring being $\sim$17 kpc away and the outer ring $\sim$10 kpc
away.  (150,+15)$^\circ$ resides in a region where both rings are present
but sparsely populated. The range of distances to the stream is $\sim$7 -
16 kpc.  (150,-15)$^\circ$ lies predominantly on their inner ring and has
a distance of $\sim$8 kpc. The detections correspond well
with this model, excluding (150,-15)$^\circ$, which as stated previously
does not resemble the expected CMD and is open to interpretation. For
a full comparison of the distances found, see
Table~\ref{DistTable}.  It is encouraging that the model provides some
support to our findings that two tidal streams are present in the (123,-19)$^\circ$ field.

\subsubsection{Non-detections: Fields (61,$\pm$15)$^\circ$, (75,$\pm$15)$^\circ$ \&
  (90,$\pm$10)$^\circ$}\label{penacomp2}

\begin{figure}\centering
\includegraphics[width=83mm]{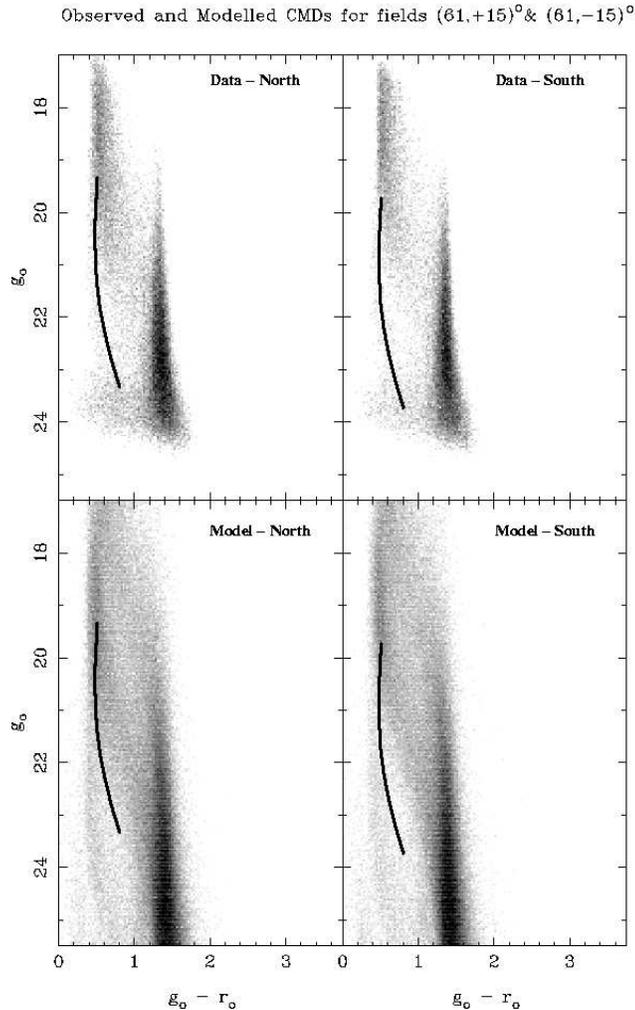}
\caption[]{Hess plots  of ({\it l},{\it b}) = (61,+15)$^\circ$ and (61,-15)$^\circ$ with the accompanying model
  CMDs. The top panels show the data obtained by the INT/WFC, visually
  we can see that the limiting magnitudes for the data here is
  $g_\circ$$\sim$23.8.  The model CMDs presented in the lower panels have
  been created with no cut in magnitude and as such continue below the limiting
  magnitudes of the data.  The main sequence overlays are those that
  show the faint Halo component trend redward brighter than the
  corresponding model plot. The associated Heliocentric distances are, $\sim$14 kpc
  for the Northern field, and $\sim$21 kpc for the Southern field.
\label{fig61pm15pena}}
\end{figure}

\begin{figure}\centering
\includegraphics[width=83mm]{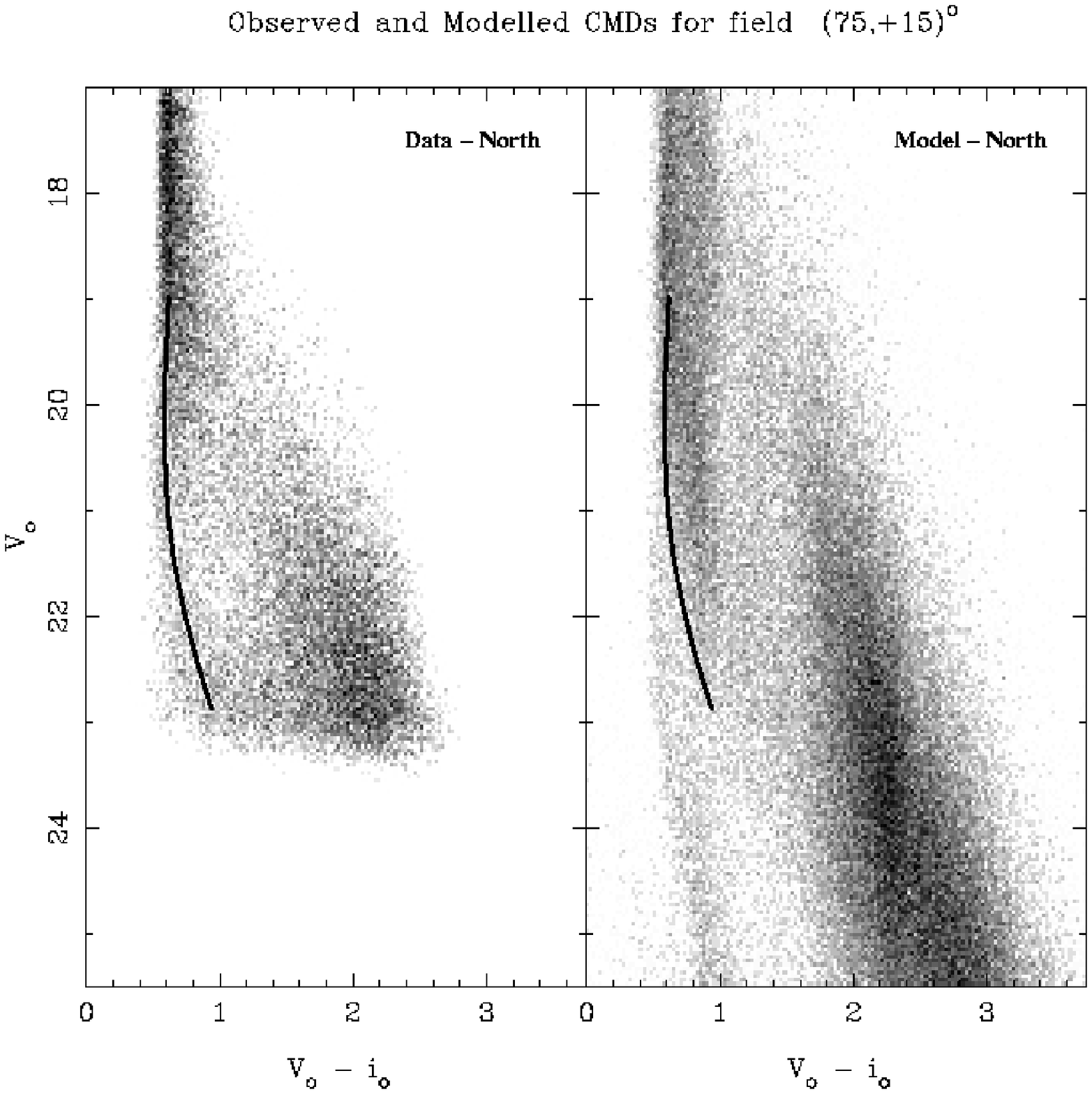}
\caption[]{Hess plots  of ({\it l},{\it b}) = (75,+15)$^\circ$ with the accompanying model
  CMDs. The top panels show the data obtained by the INT/WFC, visually
  we can see that the limiting magnitudes for the data here is
  $V_\circ$$\sim$23.2.  The model CMD presented in the right panel have
  been created with no cut in magnitude and as such continue below the limiting
  magnitudes of the data.  The main sequence overlay are those that
  show the faint Halo component turning redward brighter than the
  corresponding model plot. The associated Heliocentric distance is $\sim$15 kpc.
\label{fig75p15pena}}
\end{figure}

Considering those fields in which no ``by-eye'' detection was
made, the results of \cite{Pena2004} allow some more tentative
detections. For the (61,$\pm$15)$^\circ$ field, the Northern field stream
is located $>$14 kpc away (Heliocentric distance), in a region where
their model shows both the near and far streams crossing in the plane
of the sky [see Figure 7. of \cite{Pena2004}]. Fitting a main sequence
to the Halo structure in this field, it should be first noted that it
turns redward brighter than the model, in the same manner as the detections
in the other fields and has a distance of $\sim$16 kpc. The Southern
field, while in a region again where two streams are crossing, the
distance to these structures is $>$21 kpc and fitting a main sequence
to the Halo component here reveals a distance of $\sim$19 kpc.  The
main sequence overlays for these distance estimates can be seen in
Figure~\ref{fig61pm15pena}. The (75,$\pm$15)$^\circ$ fields,
being close to the (61,$\pm$15)$^\circ$ fields, has similar properties as
described above. The stream in the model at (75,+15)$^\circ$ has a
distance of $\sim$8 - 17 kpc. Fitting a main sequence to the Halo
component of this field returns a distance of $\sim$15 kpc, this can
be seen in Figure~\ref{fig75p15pena}. It should be noted that in
comparison with the model the location of the main sequence overlay
seems coincident with the edge of the Thick Disk (see
Figure~\ref{figmodel}).  However, since the distances match those in
the \cite{Pena2004} model it is being included in the list of
tentative new detections.  No kinematics are known about these regions
and thus are needed to resolve definitively this ambiguity. The
distance to the stream at (75,-15)$^\circ$ in the model is $\sim$20-31
kpc.  Since there is no obvious feature to fit to in our data, it can
be assumed that the stream is not present due to the distances
involved. The remaining field in which we found no ``by-eye''
signature of the stream was (90,$\pm$10)$^\circ$.  The region (90,+10)$^\circ$
is associated with the inner and outer ring of the model and has a
distance of $\sim$11 - 22 kpc.  Our observations do not support the
location of the stream in this part of the sky and no main sequence
was fitted to any part of the corresponding CMD. (90,-10)$^\circ$
resides on the edge of the outer ring which has a corresponding
distance $\sim$9 - 14 kpc and there is no obvious feature again to
fit to, possibly due to being on the edge of the stream in this
location hence a smaller number of stream stars or differences between
the model and the true location of the stream.  The subtraction figures
for the (61,$\pm$15)$^\circ$, (75,$\pm$15)$^\circ$ \&
  (90,$\pm$10)$^\circ$, as seen with the (118,+16)$^\circ$,(123,-19)$^\circ$ \&
(150,$\pm$15)$^\circ$ fields have not been shown.  Due to the relative
weakness of the structures we have identified and the slight
differences between the model and the data, the signal-to-noise
estimates of these tentative detections are not informative.

The \cite{Pena2004} model was constrained so that it accords with
the current published knowledge of the MRi detections.
The correlation with this new dataset is an encouraging sign that the
tidal stream they have modelled and the tentative detections found
here are indeed a real part of this new Milky Way structure.
 
\subsection{Could the Monoceros Ring be a detection of the Galactic warp?}
Recently, ~\citet{2004A&A...421L..29M} have claimed that the identification of
the Canis Major Dwarf (CMa) galaxy is none other than a
misinterpretation of the Galactic warp.  \citet{2004MNRAS.355L..33M} have
refuted this claim illustrating how the CMa population differs from
that of the warp. However, what certainty do we have that the MRi is
also not just a misinterpretation of the Galactic warp? 

Firstly, what signature would the warp have on the sky if
the MRi detections were merely sampling the extent of the warp?  In the synthetic
models of \citet{2003A&A...409..523R}, the Sun is assumed to be
located along a node, with the warp reaching a maximum projected angular extent
above the plane at {\it l} = 90$^\circ$ and a maximum angular extent below
the plane at {\it l} = 270$^\circ$, when viewed from the Solar
neighbourhood.  Thus all detections of the warp should be above the
plane in the region {\it l} = 0$^\circ$ - 180$^\circ$ and below the plane in the
region {\it l} = 180$^\circ$ - 360$^\circ$.  The distance to the warp would be a
minimum at each point of maximum height above the Disk and reach a maximum distance at
{\it l} = 0$^\circ$, 180$^\circ$.  Also, for any given line-of-sight the warp should
only be detected once.  Calculating the distance to the warp using the
model from ~\citet{2003A&A...409..523R}, for the lines of sight
corresponding to the INT/WFC fields, the projected distance to warp is
greater than 30 kpc.  Since this is greater than the expected extent
of the Galactic Disk, the warp should not present itself in these fields.  

In Figure~\ref{figMRidetect}, all of the detections of the MRi through
CMD identification have been plotted in a top-down view of the galaxy,
showing the coordinates of these detections.  The quavers (notes
with tails) are detections above the plane of the galaxy ({\it b} $>$
0$^\circ$) and the crochets (notes without tails) are detections
below the plane of the galaxy ({\it b} $<$ 0$^\circ$). The colours denote
the different observers.  The orange circle
marks out the solar circle around the galaxy at 8 kpc and the black
circle has a radius of 17.5 kpc.  In this plot, Galactic longitude {\it
l} is measured anti-clockwise from the line joining the Sun to the
Galactic centre.  Comparing our knowledge of how the warp signature
should present itself with Figure~\ref{figMRidetect}, it is apparent
that the detections of the MRi are not divided with those above the
plane residing solely in the region {\it l} = 0$^\circ$-180$^\circ$ and those
below at {\it l} = 180$^\circ$-360$^\circ$.  Detections of the stream have
occurred both above and below the plane at similar longitudes.  The
distances to the detections above the plane seem to lie along a great
arc of radius $\sim$17.5 kpc, which is inconsistent with the modelled
distance to the warp. This shows that the MRi is not a detection of
the warp but a distinct feature of the Milky Way.

\subsection{Conclusion}\label{conclusion}

The on-going INT/WFC Survey of the Monoceros Ring has yielded several
detections of the ring in the region {\it l,b}= (118,16)$^\circ$, (150,
15)$^\circ$ and a tentative detection at (150,-15)$^\circ$. Galactocentric distance estimates to
these structures gave $\sim$17,$\sim$17, and
$\sim$13 kpc respectively. These are combined with a reexamination of the field presented by
\citet{2003MNRAS.340L..21I}, (123,-19)$^\circ$, showing the position of
the Halo is not in accordance with the model and possibly represents
another detection of the ring. The Galactocentric distance to this feature is
estimated at $\sim$21, kpc.  This provides evidence that the ring
may be wrapped around the galaxy more than once. This is also supported by
the model of \citet{Pena2004}.

In light of claims by \citet{2004A&A...421L..29M}, that the recently
discovered Canis Major Dwarf galaxy was in fact a misinterpretation of
the Galactic warp, it seemed necessary to see whether the MRi could
also be confused with the warp.  The strongest evidence that the MRi
is not the warp is that with this survey we have shown that the MRi
can be detected on both sides of the Galactic plane at similar
Galactic longitudes.  Considering the warp cannot have this structure
the MRi can be ruled out as a result of the warp.

All the detections have been found by comparing the data with the
synthetic galactic model of \citet{2003A&A...409..523R}. This requires the assumption that the major components
of the Milky Way have smooth distributions and that major excursions
from the expected location of these features must represent something
new.  Our detections all represent major differences with the
synthetic model, typically of the order of 1 magnitude or more which
corresponds to a distance variation of $\sim$6 kpc (Heliocentrically).  The distances to the detections have been calculated
using a main-sequence style overlay modeled on the ridge line of the
SDSS S223+20 detection.  Given an offset from this original detection
we can determine a distance.  While there may be significant
systematic errors in the colour conversions, especially from ($g'$,$r'$) to
($V$,$i$) we have included the upper and lower limits of all the
distances derived to demonstrate the narrowness of the main sequences
and the intrinsic errors of this method.

Taking into account the recent paper by \cite{Pena2004}, the
observations have been re-examined and the distances and locations of
the MRi stream favourably match with their
model. Comparing the CMDs of the remaining fields reveal several more
tentative detections, namely in (61,$\pm$15)$^\circ$ \& (75,+15)$^\circ$.
These have Heliocentric distances of $\sim$16, 19 \& 15 kpc and
Galactocentric distances of $\sim$14, 17 \& 15 kpc respectively.
Each of these features had been considered earlier in the analysis of
the data, however they were deemed to weak to be significant. This
model provides support for the authenticity of these detections.
Importantly, the model of \cite{Pena2004} also confirms one of our
non-detections, (75,-15)$^\circ$ in which the distance to the stream is too
great to be detected.  The non-detections in the fields
(90,$\pm$10)$^\circ$, seemingly do not correspond to the predictions of
the model, although a deeper survey of this region may be needed to
resolve this discrepancy.  It is unknown whether the low Galactic
latitude of this field is a factor in the non-detection of the stream.

Both detections and non-detections support a complex picture of the
MRi.  In particular, those detections above the plane suggest the MRi
has an extended stream tracing an arc $\sim$17 kpc from the Galactic
centre, while the detections below the plane, reveal a tentative
detection of the TriAnd region in the background of the (123,
-19)$^\circ$ region and also the presence of a foreground stream if we
link the INT/WFC detections and those of
~\citet{2002ApJ...569..245N} \& ~\citet{2003MNRAS.340L..21I}
(Figure~\ref{figMRidetect}). Obviously, the MRi is a very complex
structure and additional observations are needed to
unravel its origins, so to this end a kinematic survey
of the region surrounding the Canis Major overdensity~\citep{2005Martin} has been conducted
with the Two Degree Field (2dF) spectrograph and so too a Wide- Camera
survey using the Anglo-Australian Observatory Wide Field Imager
(AAO/WFI) to complete the survey of the MRi. These results are still
being analysed and will form the basis of a forthcoming article.  

\begin{table*}
\begin{minipage}{185mm}
\caption{Summary of the distance estimates from all fields obtained with
  the INT/WFC, ordered in ascending Galactic longitude ({\it l}). The
  magnitude offset is from the base position of the overlay generated
  from the ridge-line in the SDSS S223+20 field
  ~\citep{2002ApJ...569..245N} followed by the associated Heliocentric
  distance and Galactocentric distance. The error associated with our
  distance estimates are all approximately $\sim$10\% of the distance, with the
  Galactocentric distance errors being of the same order as the
  Heliocentric distance errors. For comparison we have
  included rough estimates of the Heliocentric distance ranges and
  peaks of the distributions where the ~\citet{2004MNRAS.348...12M}
  and ~\citet{Pena2004} numerical simulations predict
  the tidal debris should be present.}

\begin{tabular}{||c|c|c|c|c|c|c|c|c|c||} \hline \hline
Fields\it{(l,b)$^\circ$}& Magnitude offsets &  R$_{GC}$ (kpc) & R$_{HC}$ (kpc) &
 Range R$_{HC}$\footnote{~\citet{2004MNRAS.348...12M}} (kpc)
& Peak R$_{HC}$ (kpc)&  Range R$_{HC}$\footnote{~\citet{Pena2004}} (kpc) & Peak R$_{HC}$(kpc)\\ \hline 
(61,$+$15)$^\circ$ &  +0.8      & 14         &16       & 9-11, 14-19  & 9, 16 & 14-22  & 17 \\
(61,$-$15)$^\circ$ &  +1.2      & 17         &19       & 7-10, 16-21  & 8, 18 & 21-34  & 27 \\
(75,$+$15)$^\circ$ &  +0.7      & 15         &15       & 13-20        & 18   & 8-17    & 15 \\
(75,$-$15)$^\circ$ &  -	        & -          & -       & 6-11, 14-21  & 9, 16 & 20-31  & 24 \\
(90,$+$10)$^\circ$ &  -	        & -          & -       & 5-15         & 14   & 11-22   & 18 \\
(90,$-$10)$^\circ$ &  -	        & -          & -       & 5-9, 11-24   & 7, 15 & 9-14   & 12 \\
(118,$+$16)$^\circ$& +0.2       &  17        &12       & 5-12         & 8    & 16-22   & 17 \\
(123,$-$19)$^\circ$& -0.8,+0.8  &  14,21     &8,16     & 2-13         & 10   & 7-12, 15-22 & 10, 17 \\
(150,+15)$^\circ$  & -0.4       &  17        &9        & 5-8          & 7    & 7-16    & 10 \\
(150,-15)$^\circ$  & -1.5       &  13        &6        & 12-21        & 15   & 2-12    & 8  \\
\hline\hline
\end{tabular}
\label{DistTable}
\end{minipage}
\end{table*}

\section{Acknowledgements}
BCC  would like to thank his wife, LLL, for kindly supplementing his scholarship
income, The  University of  Sydney for the University
Postgraduate Award  and the Cambridge Astronomical Survey Unit  at Cambridge
University  and  Mike  Irwin  for  their hospitality  during  my  week
there. BCC would also like to thank Jorge Pe\~{n}arrubia, for access to
his Monoceros Ring model and the anonymous referee for their many helpful
suggestions. GFL acknowledges the support of the Discovery Project
grant DP0343508.  The research of AMNF has been supported by a Marie
Curie Fellowship of the European Community under contract number
HPMF-CT-2002-01758. GFL would also like to thank  Triple J for their chillout session on
Sunday mornings which drowns out  his fighting children and also their
Three Hours of Power in case they choose to fight between the hours of
11 and 1 at night. 

\newcommand{\aap}{A\&A}
\newcommand{\apj}{ApJ}
\newcommand{\apjl}{ApJ}
\newcommand{\aaps}{AAPS}
\newcommand{\aj}{AJ}
\newcommand{\mnras}{MNRAS}
\newcommand{\nat}{Nature}

\bsp

\label{lastpage}

\end{document}